\documentclass[11pt]{article}
\usepackage{moriond,epsfig}

\def\be{\begin{equation}}
\def\ee{\end{equation}}
\def\bea{\begin{eqnarray}}
\def\eea{\end{eqnarray}}

\begin{document}
\vspace*{4cm}
\title{COMPACT OBJECTS WITH SPIN PARAMETER $a_* > 1$}

\author{COSIMO BAMBI}

\address{Institute for the Physics and Mathematics of the Universe, 
The University of Tokyo\\ Kashiwa, Chiba 277-8583, Japan}

\maketitle

\abstracts{In 4-dimensional General Relativity, black holes are 
described by the Kerr solution and are completely specified by 
their mass $M$ and by their spin angular momentum $J$. A fundamental 
limit for a black hole in General Relativity is the Kerr bound 
$|a_*| \le 1$, where $a_* = J/M^2$ is the spin parameter. Future 
experiments will be able to probe the geometry around these 
objects and test the Kerr black hole hypothesis. Interestingly, 
if these objects are not black holes, the accretion process may 
spin them up to $a_* > 1$.}

\section{Introduction}

Today we believe that the final product of the gravitational 
collapse is a black hole (BH). In 4-dimensional General Relativity, 
BHs are described by the Kerr solution and are completely specified 
by two parameters: the mass, $M$, and the spin angular momentum, 
$J$. A fundamental limit for a BH in General Relativity is the Kerr 
bound $|a_*| \le 1$, where $a_* = J/M^2$ is the spin parameter. 
For $|a_*| > 1$, the Kerr solution does not describe a BH, but a 
naked singularity, which is forbidden by the weak cosmic censorship
conjecture~\cite{wccc}.

From the observational side, we have at least two classes of
astrophysical BH candidates~\cite{ram}: stellar-mass bodies in X-ray 
binary systems ($M \sim 5 - 20$ Solar masses) and super-massive
bodies in galactic nuclei ($M \sim 10^5 - 10^{10}$ Solar masses). 
The existence of a third class of objects, intermediate-mass BH
candidates ($M \sim 10^2 - 10^4$ Solar masses), is still 
controversial, because there are not yet reliable dynamical 
measurements of their masses. All these objects are commonly 
interpreted as BHs because they cannot be explained otherwise 
without introducing new physics. The stellar-mass objects in X-ray 
binary systems are too heavy to be neutron or quark stars. At least 
some of the super-massive objects in galactic nuclei are too massive, 
compact, and old to be clusters of non-luminous bodies.

\section{Testing the Kerr Black Hole Hypothesis}

In Newtonian gravity, the potential of the gravitational field,
$\Phi$, is determined by the mass density of the matter, $\rho$,
according to the Poisson's equation, $\nabla^2 \Phi = 4 \pi G_N 
\rho$. In the exterior region, $\Phi$ can be written as 
\be\label{eq-phi}
\Phi(r,\theta,\phi) = - G_N \sum_{lm} 
\frac{{\cal M}_{lm} Y_{lm}(\theta,\phi)}{r^{l+1}} \, ,
\ee
where the coefficients ${\cal M}_{lm}$ are the multipole moments
of the gravitational field and $Y_{ml}$ are the Laplace's spherical
harmonics.

Because of the non-linear nature of the Einstein's equations,
in General Relativity it is not easy to define the counterpart
of Eq.~(\ref{eq-phi}). However, in the special case of a stationary,
axisymmetric, and asymptotically flat space-time, one can 
introduce something similar to Eq.~(\ref{eq-phi}) and define
the mass-moments ${\cal M}_n$ and the current-moments
${\cal S}_n$~\cite{hansen}. For a generic source, ${\cal M}_n$ 
and ${\cal S}_n$ are unconstrained. In the case of reflection 
symmetry, all the odd mass-moments and the even current-moments 
are identically zero. In the case of a Kerr BH, all the moments 
depend on $M$ and $J$ in a very specific way:
\be
{\cal M}_n + i {\cal S}_n = M \left(\frac{iJ}{M}\right)^n \, ,
\ee 
where $i$ is the imaginary unit; that is, $i^2 = -1$. By measuring 
the mass, the spin, and at least one more non-trivial moment of 
the gravitational field of a BH candidate (e.g the mass-quadrupole
moment $Q \equiv {\cal M}_2 = - J^2/M$), one can test the Kerr 
BH hypothesis~\cite{ryan}.

By considering the mean radiative efficiency of AGN, one can
constrain possible deviations from the Kerr geometry~\cite{agn}. 
In term of the anomalous quadrupole moment $q$, defined by 
$Q = Q_{\rm Kerr} - qM^3$, the bound is
\be
-2.00 < q < 0.14 \, .
\ee
Let us notice that this bound is already quite interesting. 
Indeed, for a self-gravitating fluid made of ordinary matter, one 
would expect $q \sim 1 - 10$. In the case of stellar-mass BH
candidates in X-ray binaries, $q$ can be potentially constrained 
by studying the soft X-ray component~\cite{bb}. 
The future detection of gravitational waves from the inspiral of 
a stellar-mass compact body into a super-massive object, the 
so-called extreme mass ratio inspiral (EMRI), will allow for 
putting much stronger constraints. LISA will be able to observe 
about $10^4 - 10^6$ gravitational wave cycles emitted by an EMRI 
while the stellar-mass body is in the strong 
field region of the super-massive object and the mass quadrupole 
moment of the latter will be measured with a precision at the level 
of $10^{-2} - 10^{-4}$~\cite{barack}.

\section{Formation of Compact Objects with $a_* > 1$}

If the current BH candidates are not the BHs predicted by General
Relativity, the Kerr bound $|a_*| \le 1$ does not hold and the
maximum value of the spin parameter may be either larger or smaller
than 1, depending on the metric around the compact object and on
its internal structure and composition. In 
Ref.~\cite{sim1,sim2,sim3,sim4}, 
I studied some features of the accretion process onto
objects with $|a_*| > 1$. However, an important question to 
address is if objects with $|a_*| > 1$ can form.

For a BH, the accretion process can spin the object up
and the final spin parameter can be very close to the Kerr bound.
In the case of a geometrically thin disk, the evolution of the 
spin parameter can be computed as follows. One assumes that the 
disk is on the equatorial plane~\footnote{For prolonged disk accretion, 
the timescale of the alignment of the spin of the object with the 
disk is much shorter than the time for the mass to increase 
significantly and it is correct to assume that the disk is on the
equatorial plane.} and that the disk's gas moves on nearly geodesic
circular orbits. The gas particles in an accretion disk fall to 
the BH by loosing energy and angular momentum. After reaching the
innermost stable circular orbit (ISCO), they are quickly swallowed 
by the BH, which changes its mass by $\delta M = \epsilon_{\rm ISCO} 
\delta m$ and its spin by $\delta J = \lambda_{\rm ISCO} \delta m$, 
where $\epsilon_{\rm ISCO}$ and $\lambda_{\rm ISCO}$ are respectively 
the specific energy and the specific angular momentum of a 
test-particle at the ISCO, while $\delta m$ is the gas rest-mass. 
The equation governing the evolution of the spin parameter is
\be\label{eq-a}
\frac{da_*}{d\ln M} = \frac{1}{M} 
\frac{\lambda_{\rm ISCO}}{\epsilon_{\rm ISCO}} - 2 a_* \, .
\ee
An initially non-rotating BH reaches the equilibrium $a_*^{eq} = 1$ 
after increasing its mass by a factor $\sqrt{6} \approx 
2.4$~\cite{bardeen}. Including the effect of the radiation emitted 
by the disk and captured by the BH, one finds $a_*^{eq} \approx 
0.998$~\cite{thorne}, because radiation with angular momentum opposite 
to the BH spin has larger capture cross section.

\begin{figure}
\par
\begin{center}
\psfig{figure=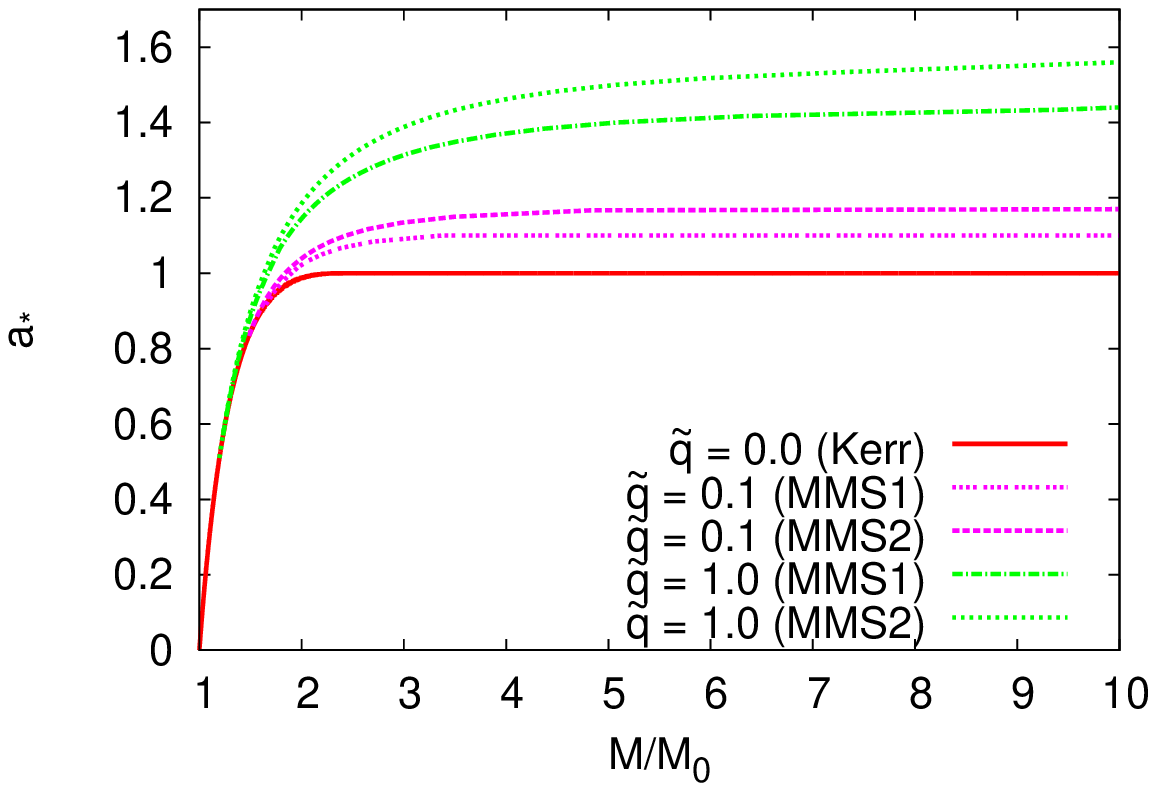,height=2.0in}
\psfig{figure=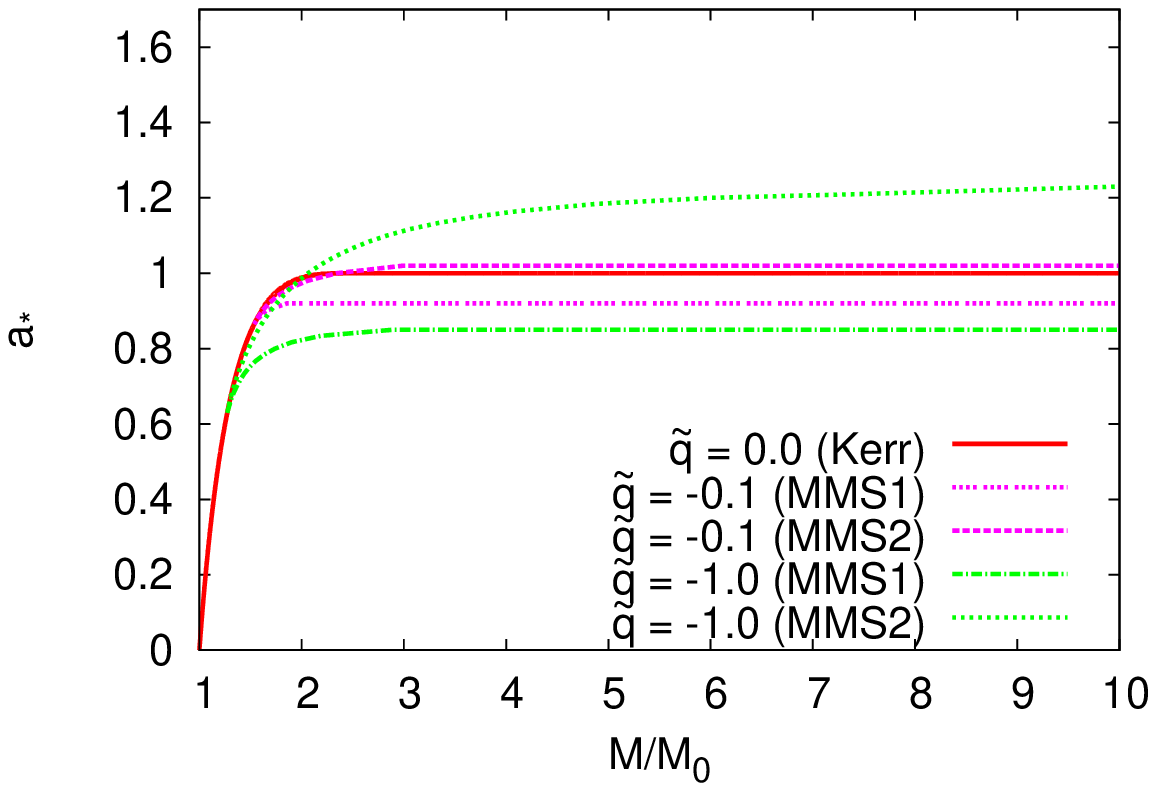,height=2.0in}
\end{center}
\par
\vspace{-5mm} 
\caption{Evolution of the spin parameter $a_*$ for an initially
non-rotating object as a function of $M/M_0$, where $M_0$ is the
mass at $a_* = 0$.
\label{fig}}
\end{figure}

As $\epsilon_{\rm ISCO}$ and $\lambda_{\rm ISCO}$ depend on the metric 
of the space-time, if the compact object is not a BH, the value
of the equilibrium spin parameter $a_*^{eq}$ may be different. 
The evolution of the spin parameter of a compact object with mass
$M$, spin angular momentum $J$, and non-Kerr quadrupole moment $Q$
was studied in~\cite{ss,ss2}. In~\cite{ss2}, I considered an extension
of the Manko-Novikov-Sanabria Go\'mez (MMS) solution~\cite{mms1,mms2},
which is a stationary, axisymmetric, and asymptotically flat exact
solution of the Einstein-Maxwell's equations. In Fig.~\ref{fig}, I
show the evolution of the spin parameter $a_*$ for different values 
of the anomalous
quadrupole moment $\tilde{q}$, defined by $Q = - (1 + \tilde{q}) 
J^2/M$. For $\tilde{q} > 0$, the compact object is more oblate than
a BH; for $\tilde{q} < 0$, the object is more prolate; for $\tilde{q} 
= 0$, one recovers exactly the Kerr metric. In Fig.~\ref{fig} there
are two curves for every value of $\tilde{q}$ because, for a given
quadrupole moment $Q$, the MMS metric may have no solutions or more
than one solution. In other words, two curves with the same $\tilde{q}$
represent the evolution of the spin parameter of two compact objects
with the same mass, spin, and mass-quadrupole moment, but different 
values of the higher order moments.

As shown in Fig.~\ref{fig}, objects more oblate than a BH ($\tilde{q}
> 0$) have an equilibrium spin parameter larger than 1. For objects
more prolate than a BH ($\tilde{q} < 0$), the situation is more
complicated, and $a_*^{eq}$ may be either larger or smaller than 1.
The origin of this fact is that for $\tilde{q} < 0$ the radius
of the ISCO may be determined by the vertical instability of the 
orbits, while for $\tilde{q} \ge 0$ (which includes Kerr BHs) 
it is always determined by the radial instability.

Lastly, let us notice that Fig.~\ref{fig} shows how, ``in principle'',
the accreting gas can spin a compact object with non-Kerr quadrupole
moment up. It may happen that the compact object becomes unstable
before reaching its natural equilibrium spin parameter. This
depends on the internal structure and composition of the object.
For example, neutron stars cannot rotate faster than about 
$\sim 1$~kHz, or $a_* \sim 0.7$. If the accretion process spins
a neutron star up above its critical value, the latter becomes
unstable and spins down by emitting gravitational waves. If the same 
thing happens to the super-massive BH candidates in galactic nuclei,
they may be an unexpected source of gravitational waves for experiments
like LISA.

\section{Conclusions}

The future gravitational wave detector LISA will be able to check 
if the super-massive objects at the center of most galaxies are
the BHs predicted by General Relativity. A fundamental limit for
a BH in General Relativity is the Kerr bound $|a_*| \le 1$, 
which is the condition for the existence of the event horizon. 
If the current BH candidates are not the BHs predicted by General
Relativity, the Kerr bound does not hold and the maximum value
of the spin parameter may be either larger or smaller than 1.
Here I showed that compact objects with $|a_*| > 1$ may form
if they have a thin disk of accretion.

\section*{Acknowledgments}

This work was supported by World Premier International Research 
Center Initiative (WPI Initiative), MEXT, Japan, and by the JSPS 
Grant-in-Aid for Young Scientists (B) No. 22740147.

\end{document}